\documentclass[pra,twocolumn]{revtex4}
\usepackage{mathtools}
\usepackage{graphicx}
\usepackage{xcolor}
\usepackage{comment}

\newcommand{\e}{\begin{equation*}\begin{aligned}}
\newcommand{\ee}{\end{aligned}\end{equation*}}

\newcommand{\en}{\begin{equation}\begin{aligned}}
\newcommand{\een}{\end{aligned} \end{equation}}

\newcommand{\f}[2]{\frac{#1}{#2}}

\newcommand{\Q}{\left}
\newcommand{\W}{\right}

\newcommand{\pma}{\begin{pmatrix}}
\newcommand{\epma}{\end{pmatrix}}

\begin{document}


\title{Time Scale for Velocity to Track a Force}

\author{C. L. Lin}
\affiliation{Physics Department, University of Houston. Houston, Texas 77024-5005, USA}

\begin{abstract}
In this paper we derive and discuss the time it takes for a force to turn a velocity. More precisely, we derive the formula for the time $\tau$  it takes a constant force that makes an angle $\alpha$ with the initial velocity $\vec{v}(0)$ to have $\vec{v}(\tau)$ get within an angle $\theta<\alpha$ of the force. We then show how the addition of a viscous force decreases $\tau$ logarithmically. The result can be generalized to any vector quantity whose first time derivative is a constant.
\end{abstract}

\maketitle

\section{Introduction}\label{section:1}

While Newton's law states that the acceleration is at all times in the direction of the force, this is not true in general of the velocity. Indeed, initially the velocity can take any direction independent of the force. However, if the force is applied long enough, the velocity will eventually approach the direction of the force, and how long it takes the force to ``turn'' this velocity is the subject of study in this paper. It should be emphasized that turning in this context means a change of direction, as opposed to a rotation, which has the additional constraint that it must preserve magnitude. \\

Although we believe this is an interesting question in its own right, a question that has a simple solution yet yields results that are rich enough to discuss and explore, we believe solving such a problem has the additional benefit that it can combat the misconception that the velocity is always in the direction of the force, a misconception that dates back to Aristotle \cite{aris}, and reinforced by the fact that most objects in our everyday experience, due to friction, are at rest relative to earth, including ourselves, so that an applied force immediately causes an object at rest to go in the direction or  ``track'' the force. We hope that by actually making a quantitative calculation of the time it takes a force to turn a velocity for the most general case, it can help dispel the notion that the velocity always tracks the force, that in general such tracking takes time to develop, and if the force changes faster than this time \footnote{The rate of change of force is proportional to the ``jerk,'' and its interpretation is discussed in \cite{jerk}.}, then the velocity needn't track force at all. We give an example of this in section \ref{section:4}. \\

\begin{figure}[h]\label{firstfigure}
\includegraphics[scale=.8]{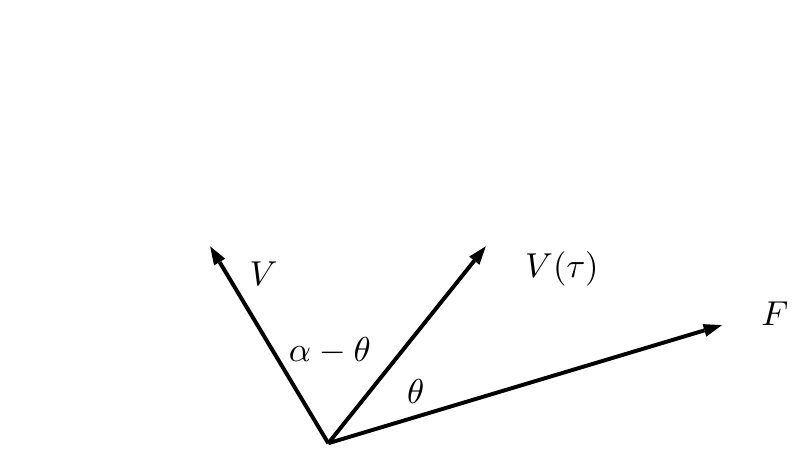} \\
\caption{$\vec{v}$, intitially at angle $\alpha$ w.r.t. $\vec{F}$, in time $\tau$ has turned through an angle $\alpha-\theta$ to be at angle $\theta$ w.r.t. $\vec{F}$. }\label{fig:1}
\end{figure}

Before we derive the result, we define our symbols and discuss what dimensional analysis can immediately tell us about the form of the answer. We write the initial velocity as $\vec{v}$ and its constant rate of change as $\vec{F}$ \footnote{$\vec{F}$ for this case would represent the force divided by the mass.}, and denote their magnitudes as $|\vec{v}|=V$ and $|\vec{F}|=F$. The initial angle between $\vec{v}$ and $\vec{F}$ is given the symbol $\alpha$. This is illustrated in Fig. \ref{firstfigure}. \\

Then from dimensional analysis, the time it takes to turn the velocity from the angle $\alpha$ to the angle $\theta<\alpha$ is given by

\en \label{eqn:1}
\tau=\f{V}{F} f(\alpha,\theta),
\een

where $f$ is a dimensionless function of dimensionless variables. The time scale is set by $\f{V}{F}$, which is the only way one can form units of time using the variables given in the problem. This factorization into a part that contains only magnitudes and a part that contains only directions is what makes the problem universal: different constant rate of change problems will differ in the ratio of quantities that sets the overall time scale, but $f(\alpha,\theta)$ would remain the same. \\


In section \ref{section:2} we will show that $\tau$ indeed has the form of Eq. \eqref{eqn:1} predicted by dimensional analysis, and also find the exact expression for $ f(\alpha,\theta)$. In section \ref{section:3}, we will look at $ f(\alpha,\theta)$ for specific cases, derive the maximum value $ f(\alpha,\theta)$ can take for a given $\theta$, and consider viscosity. Section \ref{section:4} will give some examples, followed by conclusions. 

\section{Derivation}\label{section:2}

Deriving $\tau$ in Eq. \eqref{eqn:1} is most easily done by choosing a coordinate system with the horizontal axis along $\vec{F}$. Then Fig. \ref{fig:2} shows the components of velocity at time $\tau$:

\begin{figure}[h]
\includegraphics[scale=1.2]{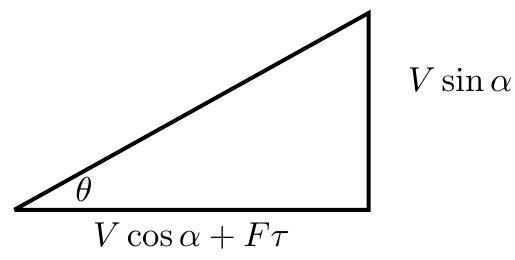} \\
\caption{The evolution of $\vec{v}$ with time. Only the component of $\vec{v}$ in the direction of $\vec{F}$, which is to the right in the picture, grows with time.}\label{fig:2}
\end{figure}

One gets:
\en \label{eqn:2}
\tan \theta&=\f{V \sin\alpha}{V\cos\alpha+F\tau}\\
\tau&=\f{V}{F} \Q(\f{\sin\alpha}{\tan\theta}-  \cos\alpha \W)=\f{V}{F}  f(\alpha,\theta).
\een

We will comment more about the $ f(\alpha,\theta)$ we derived in the next section. For now, let us point out the nature of the singularity as $\theta \rightarrow 0$ when $\alpha \neq 0,180$. This is a result of the fact that $\vec{F}$ can only change the $V \cos\alpha$ component with time, while the $V \sin\alpha$ cannot change in time, as indicated in Fig. \ref{fig:2}. Therefore a small $\theta$ requires a large lapse in time so that $V \sin\alpha$ can be made small compared to $V\cos\alpha+F\tau$, and $\theta=0$ can never be reached in a finite amount of time as the velocity will always retain the vertical component $V \sin\alpha$. This result suggests that if one wants a particle to go in a certain direction as quickly as possible, rather than just pushing in that direction, one needs to eventually push perpendicular to that direction to remove the perpendicular component of the velocity, which $\vec{F}$ is unable to do by itself.

\section{Special Cases}\label{section:3}

We now analyze Eq. \eqref{eqn:2} for some special cases. \\

\subsection{One Dimensional Motion}

For one-dimensional motion $\theta=0$ and $\alpha=180$. We get:

\en
\tau=\f{V}{F}. 
\een 

Therefore the time-scale $\f{V}{F} $ can be interpreted as the time necessary to turn a velocity 180 degrees if the force were in the opposite direction to the initial velocity. Therefore we can view $ f(\alpha,\theta)$ in $\tau=\f{V}{F} f(\alpha,\theta)$ as a correction or multiplication factor to this special case.\\

\subsection{Two Dimensional Motion}

\begin{figure}[h]
\includegraphics[scale=.8]{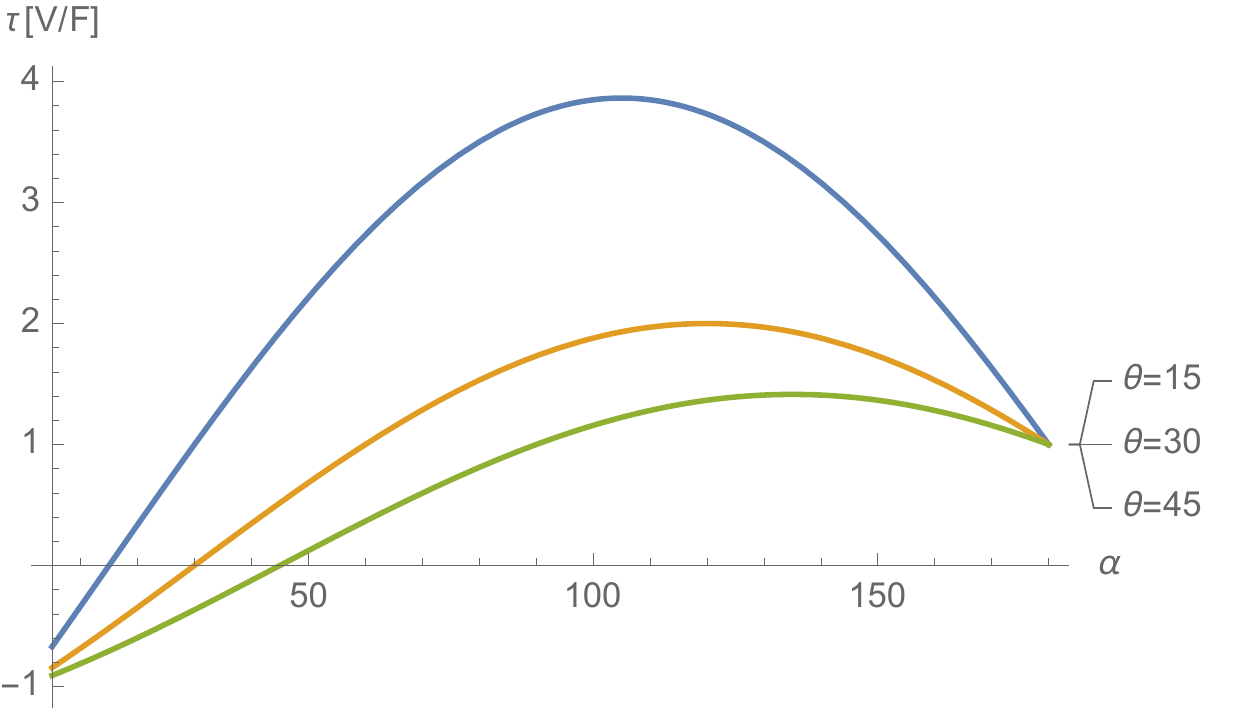}
\caption{A plot of $\tau$ vs initial angle $\alpha$ for several values of $\theta$. The curves get monotonically higher with decreasing $\theta$. $\tau$ is expressed in units of $\f{V}{F}$, or in other words, the numbers on the vertical axis represent $f(\alpha,\theta)$. }\label{fig:3}
\end{figure}

The plot of Eq. \eqref{eqn:2} for several values of $\theta$ is given in Fig. \ref{fig:3}:\\

For each value of $\theta$, in units of $\f{V}{F}$, $\tau=1$ when $\alpha=180$, which agrees with the one-dimensional case, and $\tau=0$ when $\theta=\alpha$, which agrees with the initial condition. The negative values of $\tau$ correspond to $\alpha<\theta$ which we reject, although they can be interpreted as the time in the past when they were turned to $\theta$, while further evolution in time turned them to $\alpha$ at $t=0$. For each $\theta$, the maximum time occurs when the initial angle $\alpha=90+\theta$ \footnote{This can be found by differentiating Eq. \eqref{eqn:2} w.r.t. $\alpha$ and setting to zero to find the critical point. A useful identity is $\cot \theta=\tan(\pi/2-\theta)$.}. The limiting case $\theta \rightarrow 0$ has $\tau$ peaking at 90 degrees and the curve would spike to infinity at that point, as discussed in section \ref{section:2}.\\

Since for a given $\theta$, the maximum value of $\tau$ occurs at $\alpha=90+\theta$, one can make the statement that the maximum time necessary to turn a velocity is

\en \label{eqn:numberFour}
\tau_{\text{max}}&=\f{V}{F}f(90+\theta,\theta)\\
&=\f{V}{F}\f{1}{\sin \theta},
\een

a formula that gives the peaks in Fig. \ref{fig:3}.\\
 
As an easy to remember rule of thumb, to get within an angle $\theta=30$ of the force, the most you would have to wait is $\tau=2\f{V}{F}$, or twice the one-dimensional time. This follows from plugging $\theta=30$ into Eq. \eqref{eqn:numberFour}. \\

\subsection{Addition of a Viscous Force}

The differential equation

\en \label{equationWithFriction}
\vec{F}-\eta \,\vec{v}=\f{d\vec{v}}{dt}
\een

has solution

\en 
\vec{v}(\tau)&=\f{\vec{F}}{\eta}+e^{-\eta \tau}\Q(\vec{v}_i-\f{\vec{F}}{\eta}\W)\\
&=\Q(1-e^{-\eta \tau}\W)\f{\vec{F}}{\eta}+e^{-\eta \tau}\vec{v}_{i \parallel}+e^{-\eta \tau} \vec{v}_{i \perp},
\een

where $\f{\vec{F}}{\eta}$ is the terminal velocity, and $\vec{v}_{i \parallel}$ and $\vec{v}_{i \perp}$ are the components of the initial velocity projected parallel and perpendicular to $\vec{F}$, respectively. $-\eta \vec{v}$ is the viscous drag force per mass. Performing the same analysis as with the nonviscous case:

\en \label{fric}
\tan \theta&=\f{e^{-\eta \tau} V \sin\alpha}{e^{-\eta \tau} V\cos\alpha+\Q(1-e^{-\eta \tau}\W)\f{F}{\eta}}\\
&=\f{V \sin\alpha}{ V\cos\alpha+F\Q(\f{e^{\eta \tau}-1}{\eta}\W)},
\een

which is the same as the top line of Eq. \eqref{eqn:2} but with $\tau$ replaced by $ \f{e^{\eta \tau}-1}{\eta}$. Therefore the time is $\tau=\f{1}{\eta}\ln\Q(1+\eta \tau_0\W)$ which suffers a logarithmic decrease from the nonviscous case whose time we call $\tau_0$. Since $\f{1}{\eta}\ln\Q(1+\eta \tau_0\W)\leq \tau_0$ for all $\tau_0$, with equality occurring only when  $\eta \rightarrow 0$, one can see the addition of a nonzero viscous term always makes velocity track better with force\\

\section{Examples}\label{section:4}

Consider an object, initially at rest, subjected to a forward force, followed by a backward force, followed by a forward force, followed by backward force, etc. More specifically,

\en \label{forceEqn}
F(t) = \begin{cases} F_{\text{max}} &\mbox{if } 0<t <T \\
-F_{\text{max}}  & \mbox{if }  T<t <2T\\
F_{\text{max}} &\mbox{if } 2T<t <3T \\
-F_{\text{max}} &\mbox{if } 3T<t <4T. \\
\end{cases} 
\een

\begin{figure}[h]
\includegraphics[scale=.8]{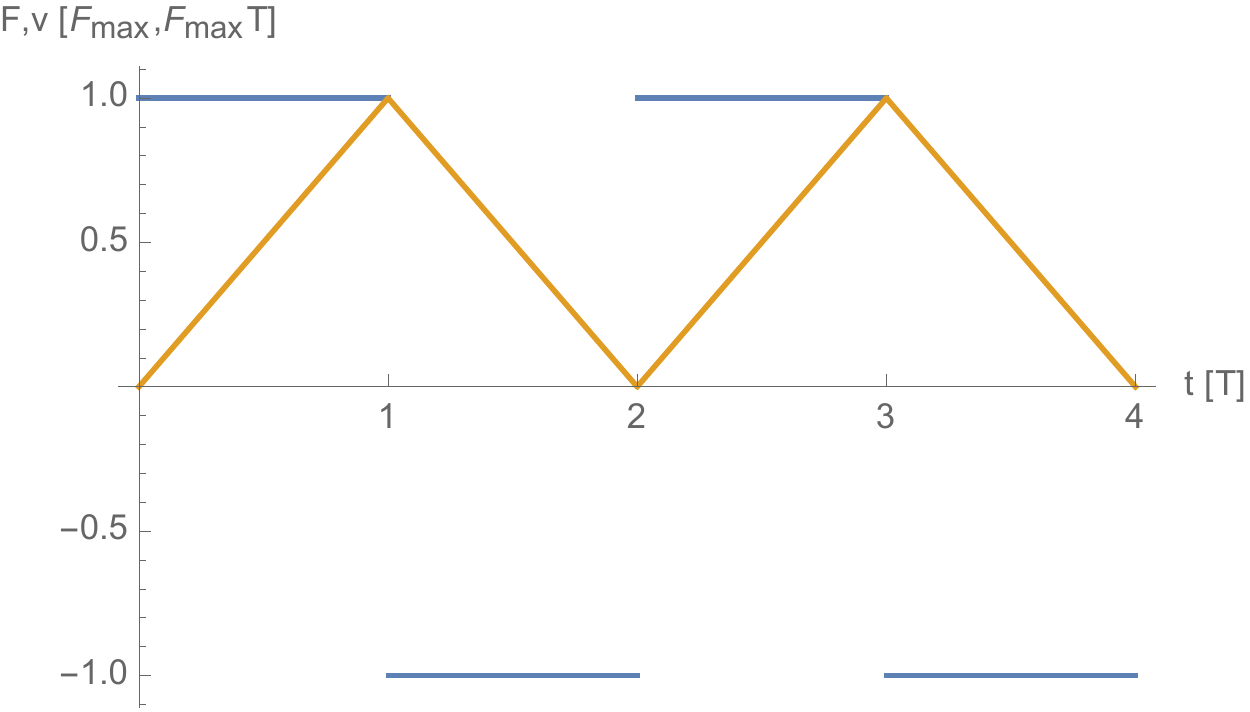}
\caption{A plot of the force and velocity as a function of the time for two cycles, for an object initially at rest. The units are expressed in terms of the magnitude of the force $F_{\text{max}}$, and the period of half a cycle, $T$. }\label{fig:4}
\end{figure}

One can see from a plot of the solution in Fig. \ref{fig:4} that the motion is always forwards, in spite of the backward push being equal in magnitude to the forwards push, and lasting the same duration. After the first time interval $T$ in which the forward force acts, the backwards force requires a time $T$ to turn the velocity backwards. However, just as this time is reached, the force changes direction to forward again. As a result, the velocity never dips below the axis (it stops but doesn't turn).\\

Such a model can be a first approximation to walking. If one is to walk on average at constant velocity, then the time-averaged acceleration and therefore the time-averaged force must be zero in a cycle: when the front foot lands friction slows a person down \cite{running}, which loses the momentum gained during the power stroke when friction acted in the forward direction. The fact that one can move with alternating forward and backward forces shows that velocity does not track force very well when walking: when the front foot lands, one does not immediately move backwards with negative velocity - instead, one slows down, just like braking a car doesn't send one backwards immediately. Indeed, walking is like start-stop traffic, which is why legs are less efficient than wheels.\\

\begin{figure}[h]
\includegraphics[scale=.8]{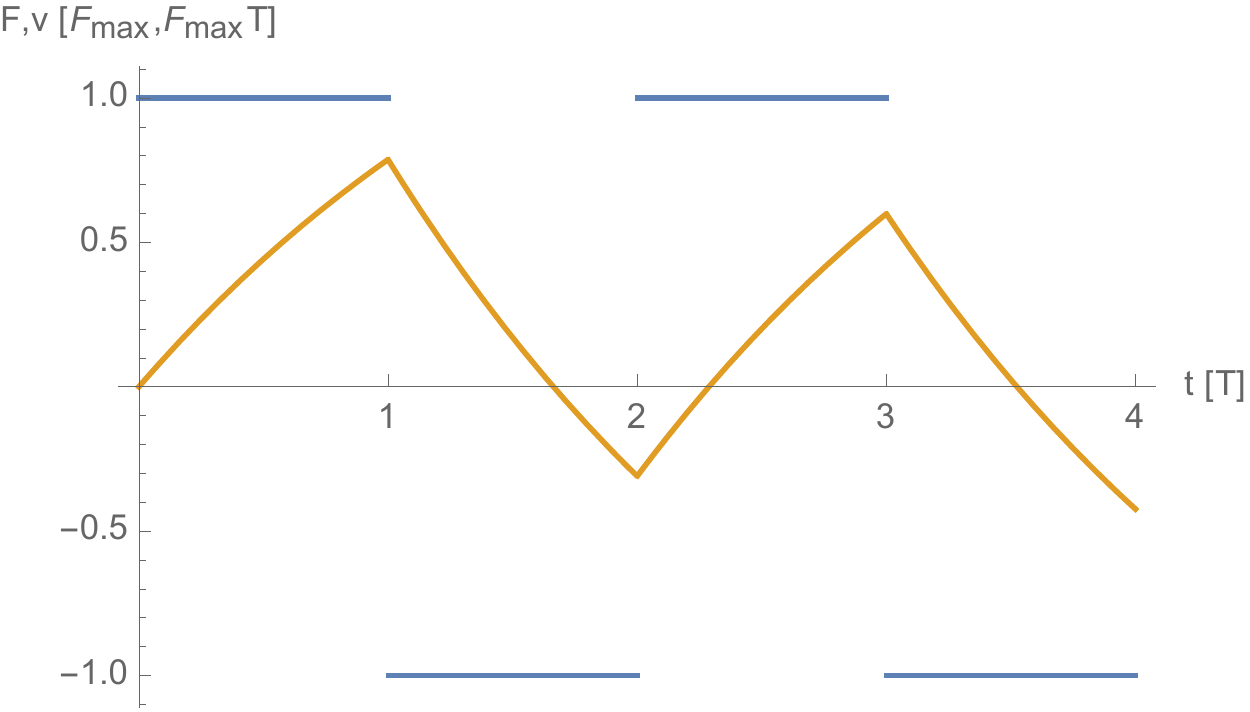}
\caption{The same plot as Fig. \ref{fig:4}, but with the addition of a viscous force with $\eta=1/2 \, T^{-1}$.}
\label{fig:5}
\end{figure}

Consider now adding a viscous force with $\eta=1/2$ (in units of $1/T$), where $\eta$ is defined in Eq. \eqref{equationWithFriction}. The solution is plotted in Fig. \ref{fig:5}.
The viscosity has allowed the velocity of the object to track the force more, and as a result, the velocity can go negative. For a lower viscosity, one can imagine a mix between Figs. \ref{fig:4} and \ref{fig:5}.\\

\begin{figure}[h]
\includegraphics[scale=.8]{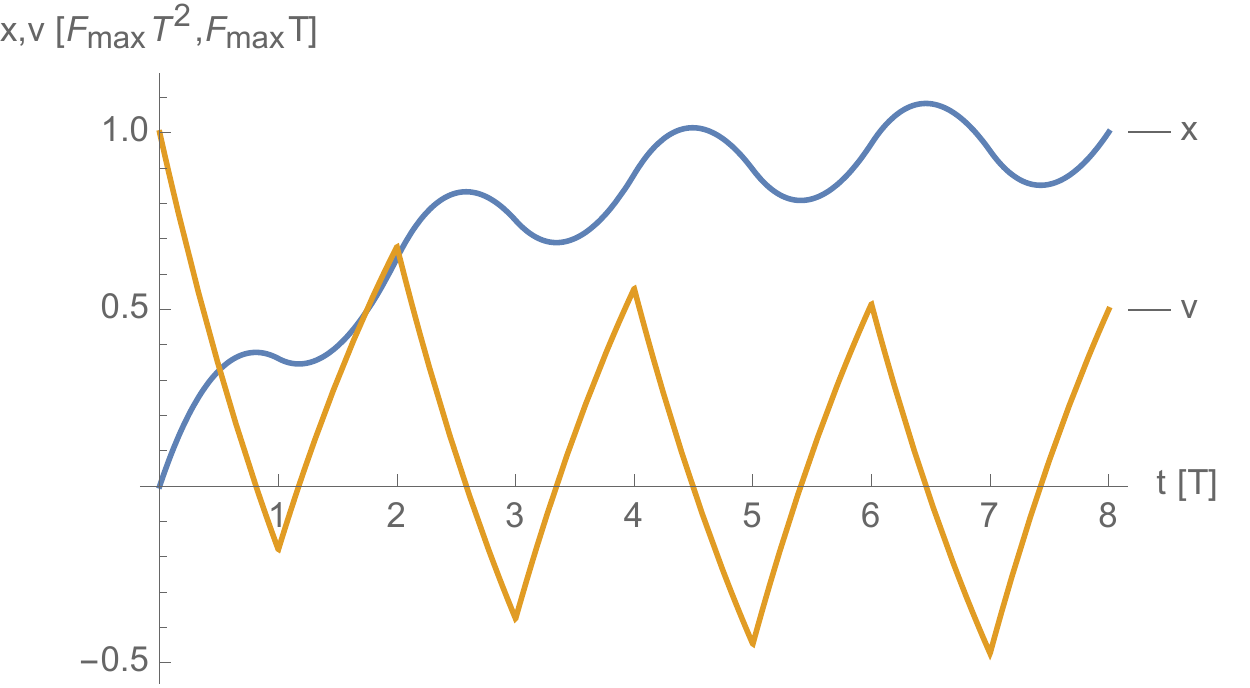}
\caption{The initial velocity has been set to $F_{\text{max}}T$, and the cycle now starts on $-F_{\text{max}}$. Position and velocity are now plotted with time.}\label{fig:6}
\end{figure}

We also plot the position and velocity for the case where $v(0)=1$ with the force starting on the negative cycle in Fig. \ref{fig:6}.
The shorter turning time has allowed the backward force to turn the object within the first half cycle $T$. Overall, one can see that as time increases, the object stagnates, unlike Fig. \ref{fig:4} where position always increases by the stride length (area under the triangle) with time. \\

The implication for swimming is that since velocity tracks force more, one has to be more careful about parts of the cycle when one is pushing the medium forward (and therefore the medium imparts a backward force) in order to reset oneself for the power stroke. The two forces need to be unequal: this can be achieved by recovering the arm above water as in the freestyle stroke, thereby making the backwards force less than if one recovered the arm underwater, or for the case of breaststroke, moving one's arms towards the centerline when underwater in order to reset for the sweeping power stroke. \\

These are just qualitative discussions as our formulas only apply to constant forces, but we hope it can serve as a starting point in a zeroth order approximation. A full, correct analysis of swimming can be found in \cite{purcell}.\\

\section{Conclusions}

In this paper we have proposed a two dimensional problem that is relatively simple to solve, yet whose solution is sufficiently rich that students can explore and interpret limiting cases. As an example, students might initially believe the turning time is proportional to $\alpha-\theta$, or the amount of turning, when in actuality it depends on those angles separately. Nevertheless, one can define a maximum turning time that is independent of initial angle $\alpha$, which is Eq. \eqref{eqn:numberFour}. \\


As an added benefit, we feel this problem emphasizes that force per mass is the rate of change of velocity, which implies that it takes time for the velocity to actually change. While students often understand this, if one dresses up the problem sufficiently, they can forget this fact: students may think that equal forwards and backwards pushes in sequence results in no net motion, yet this type of model can serve as a starting kinematic model for how we walk. Or students can think a roller coaster while upside down at the top of a vertical loop will crash to the ground, which would be true if gravity and the normal force could turn the velocity instantaneously. Of course, all these examples don't require studying this problem to solve. Nevertheless, we feel that the ``time required to turn'' can provide additional  vocabulary to describe these situations, so that the act of deriving Eq. \eqref{eqn:2} and Eq. \eqref{fric}, even if they are not used to quantitatively solve examples, can serve as a concrete basis for qualitatively talking about them.

\section{Acknowledgements}

The author would like to acknowledge the anonymous reviewers and editor for their helpful suggestions, and also Houston Community College for giving the author the opportunity to hone his teaching by allowing him to teach a few of their courses. 


\end{document}